\documentclass[envcountsect]{llncs}

\usepackage[sumlimits]{amsmath}
\usepackage{emlines}
\usepackage{emlines2}

\begin{document}

\title{The class of languages recognizable by 1-way quantum
finite automata is not closed under union
\protect\footnote{This paper also available in Latvian - contact author}}

\author{M\=aris Valdats}

\def \lket {\left|}
\def \rket {\right\rangle}
\newcommand{\ket}[1]{\lket #1\rket}
\newcommand{\comment}[1]{}
\def \cin {\!\!\in\!\!}
\def \cnotin {\!\!\notin\!\!}
\def \Qspace {l_2(Q)}

\institute{
 Institute of Mathematics and Computer Science,
 University of Latvia, Rai\c na bulv. 29, R\=\i ga,
Latvia\thanks{%
Research supported by Grant No.96.0282 from the
Latvian Council of Science}\\
 \email{sd70066@lanet.lv}
}

\maketitle

\begin{abstract}
In this paper we develop little further
the theory of quantum finite automata (QFA).
There are already few properties of QFA known,
that deterministic and probabilistic finite automata
do not have e.g. they cannot recognize all regular
languages. In this paper we show, that class
of languages recognizable by QFA is not closed
under union, even not under any Boolean operation,
where both arguments are significant.
\end{abstract}

\section{Introduction}
In recent years quantum computing is developing very quickly.
Almost all classical computational models already
have their quantum analogues.
Quantum finite automata is probably the simplest of them
and this paper is about them.
Here we will not repeat basic facts, but
as an introduction to quantum finite automata (QFA) would recommend
you these papers:
\cite{CM 97,AF 98}.
There are a lot of explanations and even examples.
Here we will recall only the definition and main results so far.

\begin{subsection}{Definition}

\begin{definition}
\label{def1}
A QFA is a tuple
$M=(Q;\Sigma ;V ;q_{0};Q_{acc};Q_{rej})$ where $Q$ is a finite set
of states, $\Sigma $ is an input alphabet, $V$ is a transition function,
$q_{0}\cin Q$ is a starting state, and $Q_{acc}\subseteq Q$
and $Q_{rej}\subseteq Q$
are sets of accepting and rejecting states
($Q_{acc}\cap Q_{rej}=\emptyset$).
The states in $Q_{acc}$ and $Q_{rej}$,
are called {\em halting states} and
the states in $Q_{non}=Q-(Q_{acc}\cup Q_{rej})$ are called
{\em non halting states}.
$\kappa$ and $\$$ are symbols that do not belong to $\Sigma$.
We use $\kappa$ and $\$$ as the left and the right endmarker,
respectively. The {\em working alphabet} of
$M$ is $\Gamma = \Sigma \cup \{\kappa ;\$\}$.

The transition function $V$ is a mapping from $\Gamma\times \Qspace$
to $\Qspace$ such that, for every $a\cin\Gamma$, the function
$V_a:\Qspace\rightarrow\Qspace$ defined by $V_a(x)=V(a, x)$ is a 
unitary transformation.
\end{definition}

The computation of a QFA starts in the superposition $|q_{0}\rangle$.
Then transformations corresponding to the left endmarker $\kappa$,
the letters of the input word $x$ and the right endmarker $\$$ are
applied. The transformation corresponding to $a\cin \Gamma$ consists
of two steps.

1. First, $V_{a}$ is applied. The new superposition $\psi^{\prime}$
is $V_{a}(\psi)$ where $\psi$ is the superposition before this step.

2. Then, $\psi^{\prime}$ is observed with respect to 
$E_{acc}, E_{rej}, E_{non}$ where
$E_{acc}=span\{|q\rangle:q\cin Q_{acc}\}$,
$E_{rej}=span\{|q\rangle :q\cin Q_{rej}\}$,
$E_{non}=span\{|q\rangle :q\cin Q_{non}\}$.
It means, that if the system's state before measurement was
$$\psi' = \sum_{q_i\in Q_{acc}} \alpha_i \ket{q_i} +
\sum_{q_j\in Q_{rej}} \beta_j \ket{q_j} +
\sum_{q_k\in Q_{non}} \gamma_k \ket{q_k}$$
then measurement accepts $\psi'$ with probability $\Sigma\alpha_i^2$,
rejects with probability $\Sigma\beta_j^2$ and
continues process with probability $\Sigma\gamma_k^2$
with system having state $\psi=\Sigma\gamma_k\ket{q_k}$.

We regard these two transformations as reading a letter $a$.
We use $V'_a$ to denote the transformation consisting of
$V_a$ followed by projection to $E_{non}$.
This is the transformation mapping $\psi$ to the non-halting part
of $V_a(\psi)$. 
We use $V_w'$ to denote the product of transformations
$V_w'=V_{a_n}'V_{a_{n-1}}'\dots V_{a_2}'V_{a_1}'$,
where $a_i$ is the $i$-th letter of the word $w$.
Also we use $\psi_y$ to denote the non-halting part of 
QFA's state after reading the left endmarker $\kappa$ and the
word $y\cin\Sigma^*$.
From the notation follows, that $\psi_w=V_{\kappa w}'(\ket{q_0})$.

We will say, that automaton recognizes language $L$ with probability $p$
$(p>\frac{1}{2})$ if automaton accepts any word $x\cin L$ with
probability $\geq p$ and
rejects any word $x\cnotin L$ with probability $\geq p$.

\end{subsection}

\begin{subsection}{Main results so far}

It has been shown \cite{KW 97}, that class of languages, recognizable
by QFA is a proper subset of regular languages.
Also it has been shown (Theorems \ref{1} and \ref{2}
taken from \cite{ABFK 99}) ,
that classes of languages recognizable
by QFA with different probabilities differs.

\begin{theorem}
\label{T11}
Let's denote hierarchy of languages
$L_n=a_1^*a_2^*a_3^*a_4^*...a_n^*$.
Then language $L_n$ can be recognized
with probability greater than $\frac{1}{2}+\frac{1}{4n}$
but not with greater than $\frac{1}{2}+\frac{3}{\sqrt{n-1}}$.
\end{theorem}

\begin{theorem}
\label{T12}
Let $L$ be a language and $M$ be its minimal automaton.
Assume that there is a word $x$ such that $M$  
contains states $q_1$, $q_2$ satisfying:
\begin{enumerate}
\item
$q_1\neq q_2$,
\item
If $M$ starts in the state $q_1$ and reads $x$,
it passes to $q_2$,
\item
If $M$ starts in the state $q_2$ and reads $x$,
it passes to $q_2$, and
\item
$q_2$ is neither "all-accepting" state, nor "all-rejecting" state.

Then $L$ cannot be recognized by a 1-way quantum finite automaton with
probability $7/9+\varepsilon$ for any fixed $\varepsilon>0$.

If we add one more condition
\item
There is a word $y$ such that if M starts in $q_2$ and reads y, it passes to $q_1$,

then $L$ cannot be recognized by any 1-way quantum finite automaton.
\end{enumerate}
\end{theorem}

%***************************************************
\hspace{0.25\textwidth}
\begin{figure}[htb]
  \centering
  \begin{Large}
%TexCad Options
%\grade{\on}
%\emlines{\on}
%\beziermacro{\off}
%\reduce{\on}
%\snapping{\off}
%\quality{2.00}
%\graddiff{0.01}
%\snapasp{1}
%\zoom{1.00}
\special{em:linewidth 0.4pt}
\unitlength 1mm
\linethickness{0.4pt}
\begin{picture}(70.00,17.00)
\put(20.00,10.00){\circle{10.00}}
\put(50.00,10.00){\circle{10.00}}
\put(20.00,10.00){\makebox(0,0)[cc]{$q_1$}}
\put(50.00,10.00){\makebox(0,0)[cc]{$q_2$}}
%\bezvec{132}(54.00,13.00)(70.00,10.00)(54.00,7.00)
\put(54.00,7.00){\vector(-4,-1){0.2}}
\emline{54.00}{13.00}{1}{56.24}{12.55}{2}
\emline{56.24}{12.55}{3}{58.11}{12.09}{4}
\emline{58.11}{12.09}{5}{59.62}{11.64}{6}
\emline{59.62}{11.64}{7}{60.76}{11.18}{8}
\emline{60.76}{11.18}{9}{61.53}{10.73}{10}
\emline{61.53}{10.73}{11}{61.93}{10.27}{12}
\emline{61.93}{10.27}{13}{61.97}{9.82}{14}
\emline{61.97}{9.82}{15}{61.64}{9.36}{16}
\emline{61.64}{9.36}{17}{60.94}{8.91}{18}
\emline{60.94}{8.91}{19}{59.88}{8.45}{20}
\emline{59.88}{8.45}{21}{58.44}{8.00}{22}
\emline{58.44}{8.00}{23}{56.64}{7.55}{24}
\emline{56.64}{7.55}{25}{54.00}{7.00}{26}
%\end
%\bezvec{92}(24.00,7.00)(35.00,3.00)(46.00,7.00)
\put(46.00,7.00){\vector(3,1){0.2}}
\emline{24.00}{7.00}{27}{26.39}{6.22}{28}
\emline{26.39}{6.22}{29}{28.78}{5.64}{30}
\emline{28.78}{5.64}{31}{31.17}{5.24}{32}
\emline{31.17}{5.24}{33}{33.57}{5.03}{34}
\emline{33.57}{5.03}{35}{35.96}{5.02}{36}
\emline{35.96}{5.02}{37}{38.35}{5.19}{38}
\emline{38.35}{5.19}{39}{40.74}{5.54}{40}
\emline{40.74}{5.54}{41}{43.13}{6.09}{42}
\emline{43.13}{6.09}{43}{46.00}{7.00}{44}
%\end
%\bezvec{92}(46.00,13.00)(35.00,17.00)(24.00,13.00)
\put(24.00,13.00){\vector(-3,-1){0.2}}
\emline{46.00}{13.00}{45}{43.61}{13.78}{46}
%\emline{43.61}{13.78}{47}{41.22}{14.36}{48}
\emline{41.22}{14.36}{49}{38.83}{14.76}{50}
%\emline{38.83}{14.76}{51}{36.43}{14.97}{52}
\emline{36.43}{14.97}{53}{34.04}{14.98}{54}
%\emline{34.04}{14.98}{55}{31.65}{14.81}{56}
\emline{31.65}{14.81}{57}{29.26}{14.46}{58}
%\emline{29.26}{14.46}{59}{26.87}{13.91}{60}
\emline{26.87}{13.91}{61}{24.00}{13.00}{62}
%\end
\put(35.00,7.00){\makebox(0,0)[cc]{$x$}}
\put(35.00,17.00){\makebox(0,0)[cc]{$y$}}
\put(58.00,14.00){\makebox(0,0)[cc]{$x$}}
\end{picture}
  \end{Large}
  \caption{Conditions of theorem \ref{T12},
  condition 5 - with dotted line}
  \label{Bilde1}
\end{figure}
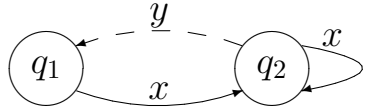
\hspace{0.25\textwidth}
%***************************************************

Theorem \ref{T11} is proved in \cite{ABFK 99},
theorem \ref{T12} is proved in \cite{AF 98}

All recently known regular languages that are not recognizable by QFA
have these properties 1-5.
The first thing we will do in next chapter, is construct a language,
that is not recognizable by a QFA, and has not the property 5.

There are also a lot of results \cite{AF 98,K 98} about number
of states needed for a QFA to recognize different languages.
It can be exponentially less than even
for probabilistic
automata but for reversible automata
(a special type of quantum automata) it can
be also exponentially more than for deterministic
automata.

It is yet unknown, what is the class of languages, recognizable by QFA.
\end{subsection}

\section{Main results}

Let's define a language $L_1=a^*bb^*a(b^*ab^*a)b^*+a^*$.
Its minimal automaton $G_1$ is

%***************************************************
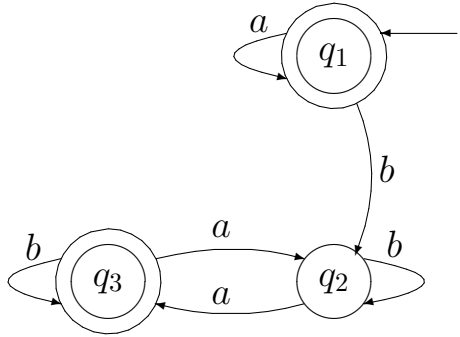
\begin{figure}[htb]
  \centering
  \begin{Large}
%TexCad Options
%\grade{\on}
%\emlines{\on}
%\beziermacro{\off}
%\reduce{\on}
%\snapping{\off}
%\quality{2.00}
%\graddiff{0.01}
%\snapasp{1}
%\zoom{1.00}
\special{em:linewidth 0.4pt}
\unitlength 1mm
\linethickness{0.4pt}
\begin{picture}(70.00,47.00)
%\circle(50.00,40.00){14.00}
\emline{50.00}{47.00}{1}{51.59}{46.82}{2}
\emline{51.59}{46.82}{3}{53.09}{46.28}{4}
\emline{53.09}{46.28}{5}{54.43}{45.42}{6}
\emline{54.43}{45.42}{7}{55.54}{44.27}{8}
\emline{55.54}{44.27}{9}{56.37}{42.91}{10}
\emline{56.37}{42.91}{11}{56.86}{41.39}{12}
\emline{56.86}{41.39}{13}{57.00}{39.80}{14}
\emline{57.00}{39.80}{15}{56.77}{38.22}{16}
\emline{56.77}{38.22}{17}{56.19}{36.73}{18}
\emline{56.19}{36.73}{19}{55.29}{35.42}{20}
\emline{55.29}{35.42}{21}{54.11}{34.34}{22}
\emline{54.11}{34.34}{23}{52.72}{33.55}{24}
\emline{52.72}{33.55}{25}{51.19}{33.10}{26}
\emline{51.19}{33.10}{27}{49.60}{33.01}{28}
\emline{49.60}{33.01}{29}{48.03}{33.28}{30}
\emline{48.03}{33.28}{31}{46.56}{33.90}{32}
\emline{46.56}{33.90}{33}{45.27}{34.84}{34}
\emline{45.27}{34.84}{35}{44.22}{36.05}{36}
\emline{44.22}{36.05}{37}{43.48}{37.46}{38}
\emline{43.48}{37.46}{39}{43.07}{39.00}{40}
\emline{43.07}{39.00}{41}{43.03}{40.60}{42}
\emline{43.03}{40.60}{43}{43.34}{42.16}{44}
\emline{43.34}{42.16}{45}{44.01}{43.61}{46}
\emline{44.01}{43.61}{47}{44.98}{44.88}{48}
\emline{44.98}{44.88}{49}{46.22}{45.89}{50}
\emline{46.22}{45.89}{51}{47.65}{46.59}{52}
\emline{47.65}{46.59}{53}{50.00}{47.00}{54}
%\end
\put(50.00,40.00){\circle{10.00}}
\put(50.00,40.00){\makebox(0,0)[cc]{$q_1$}}
%\circle(20.00,10.00){14.00}
\emline{20.00}{17.00}{55}{21.59}{16.82}{56}
\emline{21.59}{16.82}{57}{23.09}{16.28}{58}
\emline{23.09}{16.28}{59}{24.43}{15.42}{60}
\emline{24.43}{15.42}{61}{25.54}{14.27}{62}
\emline{25.54}{14.27}{63}{26.37}{12.91}{64}
\emline{26.37}{12.91}{65}{26.86}{11.39}{66}
\emline{26.86}{11.39}{67}{27.00}{9.80}{68}
\emline{27.00}{9.80}{69}{26.77}{8.22}{70}
\emline{26.77}{8.22}{71}{26.19}{6.73}{72}
\emline{26.19}{6.73}{73}{25.29}{5.42}{74}
\emline{25.29}{5.42}{75}{24.11}{4.34}{76}
\emline{24.11}{4.34}{77}{22.72}{3.55}{78}
\emline{22.72}{3.55}{79}{21.19}{3.10}{80}
\emline{21.19}{3.10}{81}{19.60}{3.01}{82}
\emline{19.60}{3.01}{83}{18.03}{3.28}{84}
\emline{18.03}{3.28}{85}{16.56}{3.90}{86}
\emline{16.56}{3.90}{87}{15.27}{4.84}{88}
\emline{15.27}{4.84}{89}{14.22}{6.05}{90}
\emline{14.22}{6.05}{91}{13.48}{7.46}{92}
\emline{13.48}{7.46}{93}{13.07}{9.00}{94}
\emline{13.07}{9.00}{95}{13.03}{10.60}{96}
\emline{13.03}{10.60}{97}{13.34}{12.16}{98}
\emline{13.34}{12.16}{99}{14.01}{13.61}{100}
\emline{14.01}{13.61}{101}{14.98}{14.88}{102}
\emline{14.98}{14.88}{103}{16.22}{15.89}{104}
\emline{16.22}{15.89}{105}{17.65}{16.59}{106}
\emline{17.65}{16.59}{107}{20.00}{17.00}{108}
%\end
\put(20.00,10.00){\circle{10.00}}
\put(50.00,10.00){\circle{10.00}}
%\bezvec{80}(46.00,7.00)(36.33,4.33)(26.33,7.00)
\put(26.33,7.00){\vector(-4,1){0.2}}
\emline{46.00}{7.00}{109}{43.58}{6.42}{110}
\emline{43.58}{6.42}{111}{41.15}{6.00}{112}
\emline{41.15}{6.00}{113}{38.70}{5.75}{114}
\emline{38.70}{5.75}{115}{33.79}{5.75}{116}
\emline{33.79}{5.75}{117}{31.31}{6.00}{118}
\emline{31.31}{6.00}{119}{28.83}{6.42}{120}
\emline{28.83}{6.42}{121}{26.33}{7.00}{122}
%\end
%\bezvec{80}(26.33,13.00)(37.00,15.67)(46.00,13.00)
\put(46.00,13.00){\vector(4,-1){0.2}}
\emline{26.33}{13.00}{123}{28.97}{13.58}{124}
\emline{28.97}{13.58}{125}{31.56}{14.00}{126}
\emline{31.56}{14.00}{127}{34.10}{14.25}{128}
\emline{34.10}{14.25}{129}{39.02}{14.25}{130}
\emline{39.02}{14.25}{131}{41.40}{14.00}{132}
\emline{41.40}{14.00}{133}{43.72}{13.58}{134}
\emline{43.72}{13.58}{135}{46.00}{13.00}{136}
%\end
\put(20.00,10.00){\makebox(0,0)[cc]{$q_3$}}
\put(50.00,10.00){\makebox(0,0)[cc]{$q_2$}}
%\bezvec{132}(54.00,13.00)(70.00,10.00)(54.00,7.00)
\put(54.00,7.00){\vector(-4,-1){0.2}}
\emline{54.00}{13.00}{137}{56.24}{12.55}{138}
\emline{56.24}{12.55}{139}{58.11}{12.09}{140}
\emline{58.11}{12.09}{141}{59.62}{11.64}{142}
\emline{59.62}{11.64}{143}{60.76}{11.18}{144}
\emline{60.76}{11.18}{145}{61.53}{10.73}{146}
\emline{61.53}{10.73}{147}{61.93}{10.27}{148}
\emline{61.93}{10.27}{149}{61.97}{9.82}{150}
\emline{61.97}{9.82}{151}{61.64}{9.36}{152}
\emline{61.64}{9.36}{153}{60.94}{8.91}{154}
\emline{60.94}{8.91}{155}{59.88}{8.45}{156}
\emline{59.88}{8.45}{157}{58.44}{8.00}{158}
\emline{58.44}{8.00}{159}{56.64}{7.55}{160}
\emline{56.64}{7.55}{161}{54.00}{7.00}{162}
%\end
%\bezvec{116}(13.67,13.00)(-0.33,10.00)(13.67,7.00)
\put(13.67,7.00){\vector(4,-1){0.2}}
\emline{13.67}{13.00}{163}{11.46}{12.48}{164}
\emline{11.46}{12.48}{165}{9.67}{11.97}{166}
\emline{9.67}{11.97}{167}{8.30}{11.45}{168}
\emline{8.30}{11.45}{169}{7.34}{10.93}{170}
\emline{7.34}{10.93}{171}{6.80}{10.41}{172}
\emline{6.80}{10.41}{173}{6.67}{9.90}{174}
\emline{6.67}{9.90}{175}{6.97}{9.38}{176}
\emline{6.97}{9.38}{177}{7.67}{8.86}{178}
\emline{7.67}{8.86}{179}{8.80}{8.34}{180}
\emline{8.80}{8.34}{181}{10.34}{7.83}{182}
\emline{10.34}{7.83}{183}{13.67}{7.00}{184}
%\end
%\bezvec{84}(53.00,33.67)(57.00,24.00)(53.00,14.00)
\put(53.00,14.00){\vector(-1,-3){0.2}}
\emline{53.00}{33.67}{185}{53.84}{31.36}{186}
\emline{53.84}{31.36}{187}{54.45}{29.04}{188}
\emline{54.45}{29.04}{189}{54.84}{26.72}{190}
\emline{54.84}{26.72}{191}{55.00}{24.38}{192}
\emline{55.00}{24.38}{193}{54.93}{22.04}{194}
\emline{54.93}{22.04}{195}{54.63}{19.69}{196}
\emline{54.63}{19.69}{197}{54.11}{17.32}{198}
\emline{54.11}{17.32}{199}{53.00}{14.00}{200}
%\end
\put(57.00,25.00){\makebox(0,0)[cc]{$b$}}
\put(58.00,15.00){\makebox(0,0)[cc]{$b$}}
\put(35.00,17.00){\makebox(0,0)[cc]{$a$}}
\put(35.00,8.00){\makebox(0,0)[cc]{$a$}}
\put(10.00,14.67){\makebox(0,0)[cc]{$b$}}
%\vector(66.33,43.00)(56.33,43.00)
\put(56.33,43.00){\vector(-1,0){0.2}}
\emline{66.33}{43.00}{201}{56.33}{43.00}{202}
%\end
%\bezvec{116}(43.67,43.00)(29.67,40.00)(43.67,37.00)
\put(43.67,37.00){\vector(4,-1){0.2}}
\emline{43.67}{43.00}{203}{41.46}{42.48}{204}
\emline{41.46}{42.48}{205}{39.67}{41.97}{206}
\emline{39.67}{41.97}{207}{38.30}{41.45}{208}
\emline{38.30}{41.45}{209}{37.34}{40.93}{210}
\emline{37.34}{40.93}{211}{36.80}{40.41}{212}
\emline{36.80}{40.41}{213}{36.67}{39.90}{214}
\emline{36.67}{39.90}{215}{36.97}{39.38}{216}
\emline{36.97}{39.38}{217}{37.67}{38.86}{218}
\emline{37.67}{38.86}{219}{38.80}{38.34}{220}
\emline{38.80}{38.34}{221}{40.34}{37.83}{222}
\emline{40.34}{37.83}{223}{43.67}{37.00}{224}
%\end
\put(40.00,44.33){\makebox(0,0)[cc]{$a$}}
\end{picture}
  \end{Large}
  \caption{Automaton $G_1$}
  \label{Bilde2}
\end{figure}
%***************************************************

States $q_1$ and $q_2$ of automaton $G_1$ and the word $b$
fulfills conditions 1-4 of theorem \ref{T12}
but condition 5 is not fulfilled.

\begin{theorem}
\label{Tm}
Language $L_1$ is not recognizable by a QFA.
\end{theorem}

\begin{proof}
As it is long and technical, it is presented in appendix.
\end{proof}

Now let's consider 2 other languages $L_2$ and $L_3$. For variety
they will be recognizable by QFA. So they are:

$L_2=(aa)^*bb^*a(b^*ab^*a)b^*+(aa)^*$

$L_3=aL_2=a(aa)^*bb^*a(b^*ab^*a)b^*+a(aa)^*$

More easy is to look at their minimal automatons $G_2$ and $G_3$
(Fig.\ref{Bilde3} and Fig.\ref{Bilde4})
They differ only with a starting state. That is the
only thing, where their quantum analogs
$K_2$ and $K_3$ are going to differ, too.

%***************************************************
\begin{figure}[htb]
  \begin{minipage}{0.4\textwidth}
    \centering
    \begin{Large}
%TexCad Options
%\grade{\on}
%\emlines{\on}
%\beziermacro{\off}
%\reduce{\on}
%\snapping{\off}
%\quality{2.00}
%\graddiff{0.01}
%\snapasp{1}
%\zoom{1.00}
\special{em:linewidth 0.4pt}
\unitlength 1mm
\linethickness{0.4pt}
\begin{picture}(70.00,47.00)
%\circle(50.00,40.00){14.00}
\emline{50.00}{47.00}{1}{51.59}{46.82}{2}
\emline{51.59}{46.82}{3}{53.09}{46.28}{4}
\emline{53.09}{46.28}{5}{54.43}{45.42}{6}
\emline{54.43}{45.42}{7}{55.54}{44.27}{8}
\emline{55.54}{44.27}{9}{56.37}{42.91}{10}
\emline{56.37}{42.91}{11}{56.86}{41.39}{12}
\emline{56.86}{41.39}{13}{57.00}{39.80}{14}
\emline{57.00}{39.80}{15}{56.77}{38.22}{16}
\emline{56.77}{38.22}{17}{56.19}{36.73}{18}
\emline{56.19}{36.73}{19}{55.29}{35.42}{20}
\emline{55.29}{35.42}{21}{54.11}{34.34}{22}
\emline{54.11}{34.34}{23}{52.72}{33.55}{24}
\emline{52.72}{33.55}{25}{51.19}{33.10}{26}
\emline{51.19}{33.10}{27}{49.60}{33.01}{28}
\emline{49.60}{33.01}{29}{48.03}{33.28}{30}
\emline{48.03}{33.28}{31}{46.56}{33.90}{32}
\emline{46.56}{33.90}{33}{45.27}{34.84}{34}
\emline{45.27}{34.84}{35}{44.22}{36.05}{36}
\emline{44.22}{36.05}{37}{43.48}{37.46}{38}
\emline{43.48}{37.46}{39}{43.07}{39.00}{40}
\emline{43.07}{39.00}{41}{43.03}{40.60}{42}
\emline{43.03}{40.60}{43}{43.34}{42.16}{44}
\emline{43.34}{42.16}{45}{44.01}{43.61}{46}
\emline{44.01}{43.61}{47}{44.98}{44.88}{48}
\emline{44.98}{44.88}{49}{46.22}{45.89}{50}
\emline{46.22}{45.89}{51}{47.65}{46.59}{52}
\emline{47.65}{46.59}{53}{50.00}{47.00}{54}
%\end
\put(50.00,40.00){\circle{10.00}}
\put(20.00,40.00){\circle{10.00}}
%\bezvec{80}(24.00,43.00)(33.67,45.67)(43.67,43.00)
\put(43.67,43.00){\vector(4,-1){0.2}}
\emline{24.00}{43.00}{55}{26.42}{43.58}{56}
\emline{26.42}{43.58}{57}{28.85}{44.00}{58}
\emline{28.85}{44.00}{59}{31.30}{44.25}{60}
\emline{31.30}{44.25}{61}{36.21}{44.25}{62}
\emline{36.21}{44.25}{63}{38.69}{44.00}{64}
\emline{38.69}{44.00}{65}{41.17}{43.58}{66}
\emline{41.17}{43.58}{67}{43.67}{43.00}{68}
%\end
%\bezvec{80}(43.67,37.00)(33.00,34.33)(24.00,37.00)
\put(24.00,37.00){\vector(-4,1){0.2}}
\emline{43.67}{37.00}{69}{41.03}{36.42}{70}
\emline{41.03}{36.42}{71}{38.44}{36.00}{72}
\emline{38.44}{36.00}{73}{35.90}{35.75}{74}
\emline{35.90}{35.75}{75}{30.98}{35.75}{76}
\emline{30.98}{35.75}{77}{28.60}{36.00}{78}
\emline{28.60}{36.00}{79}{26.28}{36.42}{80}
\emline{26.28}{36.42}{81}{24.00}{37.00}{82}
%\end
\put(50.00,40.00){\makebox(0,0)[cc]{$q_1$}}
\put(20.00,40.00){\makebox(0,0)[cc]{$q_4$}}
%\circle(20.00,10.00){14.00}
\emline{20.00}{17.00}{83}{21.59}{16.82}{84}
\emline{21.59}{16.82}{85}{23.09}{16.28}{86}
\emline{23.09}{16.28}{87}{24.43}{15.42}{88}
\emline{24.43}{15.42}{89}{25.54}{14.27}{90}
\emline{25.54}{14.27}{91}{26.37}{12.91}{92}
\emline{26.37}{12.91}{93}{26.86}{11.39}{94}
\emline{26.86}{11.39}{95}{27.00}{9.80}{96}
\emline{27.00}{9.80}{97}{26.77}{8.22}{98}
\emline{26.77}{8.22}{99}{26.19}{6.73}{100}
\emline{26.19}{6.73}{101}{25.29}{5.42}{102}
\emline{25.29}{5.42}{103}{24.11}{4.34}{104}
\emline{24.11}{4.34}{105}{22.72}{3.55}{106}
\emline{22.72}{3.55}{107}{21.19}{3.10}{108}
\emline{21.19}{3.10}{109}{19.60}{3.01}{110}
\emline{19.60}{3.01}{111}{18.03}{3.28}{112}
\emline{18.03}{3.28}{113}{16.56}{3.90}{114}
\emline{16.56}{3.90}{115}{15.27}{4.84}{116}
\emline{15.27}{4.84}{117}{14.22}{6.05}{118}
\emline{14.22}{6.05}{119}{13.48}{7.46}{120}
\emline{13.48}{7.46}{121}{13.07}{9.00}{122}
\emline{13.07}{9.00}{123}{13.03}{10.60}{124}
\emline{13.03}{10.60}{125}{13.34}{12.16}{126}
\emline{13.34}{12.16}{127}{14.01}{13.61}{128}
\emline{14.01}{13.61}{129}{14.98}{14.88}{130}
\emline{14.98}{14.88}{131}{16.22}{15.89}{132}
\emline{16.22}{15.89}{133}{17.65}{16.59}{134}
\emline{17.65}{16.59}{135}{20.00}{17.00}{136}
%\end
\put(20.00,10.00){\circle{10.00}}
\put(50.00,10.00){\circle{10.00}}
%\bezvec{80}(46.00,7.00)(36.33,4.33)(26.33,7.00)
\put(26.33,7.00){\vector(-4,1){0.2}}
\emline{46.00}{7.00}{137}{43.58}{6.42}{138}
\emline{43.58}{6.42}{139}{41.15}{6.00}{140}
\emline{41.15}{6.00}{141}{38.70}{5.75}{142}
\emline{38.70}{5.75}{143}{33.79}{5.75}{144}
\emline{33.79}{5.75}{145}{31.31}{6.00}{146}
\emline{31.31}{6.00}{147}{28.83}{6.42}{148}
\emline{28.83}{6.42}{149}{26.33}{7.00}{150}
%\end
%\bezvec{80}(26.33,13.00)(37.00,15.67)(46.00,13.00)
\put(46.00,13.00){\vector(4,-1){0.2}}
\emline{26.33}{13.00}{151}{28.97}{13.58}{152}
\emline{28.97}{13.58}{153}{31.56}{14.00}{154}
\emline{31.56}{14.00}{155}{34.10}{14.25}{156}
\emline{34.10}{14.25}{157}{39.02}{14.25}{158}
\emline{39.02}{14.25}{159}{41.40}{14.00}{160}
\emline{41.40}{14.00}{161}{43.72}{13.58}{162}
\emline{43.72}{13.58}{163}{46.00}{13.00}{164}
%\end
\put(20.00,10.00){\makebox(0,0)[cc]{$q_3$}}
\put(50.00,10.00){\makebox(0,0)[cc]{$q_2$}}
%\bezvec{132}(54.00,13.00)(70.00,10.00)(54.00,7.00)
\put(54.00,7.00){\vector(-4,-1){0.2}}
\emline{54.00}{13.00}{165}{56.24}{12.55}{166}
\emline{56.24}{12.55}{167}{58.11}{12.09}{168}
\emline{58.11}{12.09}{169}{59.62}{11.64}{170}
\emline{59.62}{11.64}{171}{60.76}{11.18}{172}
\emline{60.76}{11.18}{173}{61.53}{10.73}{174}
\emline{61.53}{10.73}{175}{61.93}{10.27}{176}
\emline{61.93}{10.27}{177}{61.97}{9.82}{178}
\emline{61.97}{9.82}{179}{61.64}{9.36}{180}
\emline{61.64}{9.36}{181}{60.94}{8.91}{182}
\emline{60.94}{8.91}{183}{59.88}{8.45}{184}
\emline{59.88}{8.45}{185}{58.44}{8.00}{186}
\emline{58.44}{8.00}{187}{56.64}{7.55}{188}
\emline{56.64}{7.55}{189}{54.00}{7.00}{190}
%\end
%\bezvec{116}(13.67,13.00)(-0.33,10.00)(13.67,7.00)
\put(13.67,7.00){\vector(4,-1){0.2}}
\emline{13.67}{13.00}{191}{11.46}{12.48}{192}
\emline{11.46}{12.48}{193}{9.67}{11.97}{194}
\emline{9.67}{11.97}{195}{8.30}{11.45}{196}
\emline{8.30}{11.45}{197}{7.34}{10.93}{198}
\emline{7.34}{10.93}{199}{6.80}{10.41}{200}
\emline{6.80}{10.41}{201}{6.67}{9.90}{202}
\emline{6.67}{9.90}{203}{6.97}{9.38}{204}
\emline{6.97}{9.38}{205}{7.67}{8.86}{206}
\emline{7.67}{8.86}{207}{8.80}{8.34}{208}
\emline{8.80}{8.34}{209}{10.34}{7.83}{210}
\emline{10.34}{7.83}{211}{13.67}{7.00}{212}
%\end
\put(5.00,25.00){\circle{10.00}}
\put(5.00,25.00){\makebox(0,0)[cc]{$q_5$}}
%\bezvec{132}(9.00,28.00)(25.00,25.00)(9.00,22.00)
\put(9.00,22.00){\vector(-4,-1){0.2}}
\emline{9.00}{28.00}{213}{11.24}{27.55}{214}
\emline{11.24}{27.55}{215}{13.11}{27.09}{216}
\emline{13.11}{27.09}{217}{14.62}{26.64}{218}
\emline{14.62}{26.64}{219}{15.76}{26.18}{220}
\emline{15.76}{26.18}{221}{16.53}{25.73}{222}
\emline{16.53}{25.73}{223}{16.93}{25.27}{224}
\emline{16.93}{25.27}{225}{16.97}{24.82}{226}
\emline{16.97}{24.82}{227}{16.64}{24.36}{228}
\emline{16.64}{24.36}{229}{15.94}{23.91}{230}
\emline{15.94}{23.91}{231}{14.88}{23.45}{232}
\emline{14.88}{23.45}{233}{13.44}{23.00}{234}
\emline{13.44}{23.00}{235}{11.64}{22.55}{236}
\emline{11.64}{22.55}{237}{9.00}{22.00}{238}
%\end
\put(12.67,30.00){\makebox(0,0)[cc]{a,b}}
%\bezvec{84}(53.00,33.67)(57.00,24.00)(53.00,14.00)
\put(53.00,14.00){\vector(-1,-3){0.2}}
\emline{53.00}{33.67}{239}{53.84}{31.36}{240}
\emline{53.84}{31.36}{241}{54.45}{29.04}{242}
\emline{54.45}{29.04}{243}{54.84}{26.72}{244}
\emline{54.84}{26.72}{245}{55.00}{24.38}{246}
\emline{55.00}{24.38}{247}{54.93}{22.04}{248}
\emline{54.93}{22.04}{249}{54.63}{19.69}{250}
\emline{54.63}{19.69}{251}{54.11}{17.32}{252}
\emline{54.11}{17.32}{253}{53.00}{14.00}{254}
%\end
\put(6.00,40.00){\makebox(0,0)[cc]{$b$}}
\put(35.00,47.00){\makebox(0,0)[cc]{$a$}}
\put(35.00,39.00){\makebox(0,0)[cc]{$a$}}
\put(57.00,25.00){\makebox(0,0)[cc]{$b$}}
\put(58.00,15.00){\makebox(0,0)[cc]{$b$}}
\put(35.00,17.00){\makebox(0,0)[cc]{$a$}}
\put(35.00,8.00){\makebox(0,0)[cc]{$a$}}
\put(10.00,14.67){\makebox(0,0)[cc]{$b$}}
%\vector(6.00,43.00)(16.00,43.00)
\put(16.00,43.00){\vector(1,0){0.2}}
\emline{6.00}{43.00}{255}{16.00}{43.00}{256}
%\end
%\bezvec{72}(15.00,40.00)(6.00,39.00)(5.00,30.00)
\put(5.00,30.00){\vector(-1,-4){0.2}}
\emline{15.00}{40.00}{257}{12.65}{39.57}{258}
\emline{12.65}{39.57}{259}{10.62}{38.83}{260}
\emline{10.62}{38.83}{261}{8.89}{37.78}{262}
\emline{8.89}{37.78}{263}{7.47}{36.42}{264}
\emline{7.47}{36.42}{265}{6.36}{34.75}{266}
\emline{6.36}{34.75}{267}{5.56}{32.78}{268}
\emline{5.56}{32.78}{269}{5.00}{30.00}{270}
%\end
\end{picture}
    \end{Large}
    \setlength{\abovecaptionskip}{0pt}
    \caption{Automaton $G_2$}
    \label{Bilde3}
  \end{minipage}
  \hspace{0.1\textwidth}
  \begin{minipage}{0.4\textwidth}
    \centering
    \begin{Large}
%TexCad Options
%\grade{\on}
%\emlines{\on}
%\beziermacro{\off}
%\reduce{\on}
%\snapping{\off}
%\quality{2.00}
%\graddiff{0.01}
%\snapasp{1}
%\zoom{1.00}
\special{em:linewidth 0.4pt}
\unitlength 1mm
\linethickness{0.4pt}
\begin{picture}(70.00,47.00)
%\circle(50.00,40.00){14.00}
\emline{50.00}{47.00}{1}{51.59}{46.82}{2}
\emline{51.59}{46.82}{3}{53.09}{46.28}{4}
\emline{53.09}{46.28}{5}{54.43}{45.42}{6}
\emline{54.43}{45.42}{7}{55.54}{44.27}{8}
\emline{55.54}{44.27}{9}{56.37}{42.91}{10}
\emline{56.37}{42.91}{11}{56.86}{41.39}{12}
\emline{56.86}{41.39}{13}{57.00}{39.80}{14}
\emline{57.00}{39.80}{15}{56.77}{38.22}{16}
\emline{56.77}{38.22}{17}{56.19}{36.73}{18}
\emline{56.19}{36.73}{19}{55.29}{35.42}{20}
\emline{55.29}{35.42}{21}{54.11}{34.34}{22}
\emline{54.11}{34.34}{23}{52.72}{33.55}{24}
\emline{52.72}{33.55}{25}{51.19}{33.10}{26}
\emline{51.19}{33.10}{27}{49.60}{33.01}{28}
\emline{49.60}{33.01}{29}{48.03}{33.28}{30}
\emline{48.03}{33.28}{31}{46.56}{33.90}{32}
\emline{46.56}{33.90}{33}{45.27}{34.84}{34}
\emline{45.27}{34.84}{35}{44.22}{36.05}{36}
\emline{44.22}{36.05}{37}{43.48}{37.46}{38}
\emline{43.48}{37.46}{39}{43.07}{39.00}{40}
\emline{43.07}{39.00}{41}{43.03}{40.60}{42}
\emline{43.03}{40.60}{43}{43.34}{42.16}{44}
\emline{43.34}{42.16}{45}{44.01}{43.61}{46}
\emline{44.01}{43.61}{47}{44.98}{44.88}{48}
\emline{44.98}{44.88}{49}{46.22}{45.89}{50}
\emline{46.22}{45.89}{51}{47.65}{46.59}{52}
\emline{47.65}{46.59}{53}{50.00}{47.00}{54}
%\end
\put(50.00,40.00){\circle{10.00}}
\put(20.00,40.00){\circle{10.00}}
%\bezvec{80}(24.00,43.00)(33.67,45.67)(43.67,43.00)
\put(43.67,43.00){\vector(4,-1){0.2}}
\emline{24.00}{43.00}{55}{26.42}{43.58}{56}
\emline{26.42}{43.58}{57}{28.85}{44.00}{58}
\emline{28.85}{44.00}{59}{31.30}{44.25}{60}
\emline{31.30}{44.25}{61}{36.21}{44.25}{62}
\emline{36.21}{44.25}{63}{38.69}{44.00}{64}
\emline{38.69}{44.00}{65}{41.17}{43.58}{66}
\emline{41.17}{43.58}{67}{43.67}{43.00}{68}
%\end
%\bezvec{80}(43.67,37.00)(33.00,34.33)(24.00,37.00)
\put(24.00,37.00){\vector(-4,1){0.2}}
\emline{43.67}{37.00}{69}{41.03}{36.42}{70}
\emline{41.03}{36.42}{71}{38.44}{36.00}{72}
\emline{38.44}{36.00}{73}{35.90}{35.75}{74}
\emline{35.90}{35.75}{75}{30.98}{35.75}{76}
\emline{30.98}{35.75}{77}{28.60}{36.00}{78}
\emline{28.60}{36.00}{79}{26.28}{36.42}{80}
\emline{26.28}{36.42}{81}{24.00}{37.00}{82}
%\end
\put(50.00,40.00){\makebox(0,0)[cc]{$q_1$}}
\put(20.00,40.00){\makebox(0,0)[cc]{$q_4$}}
%\circle(20.00,10.00){14.00}
\emline{20.00}{17.00}{83}{21.59}{16.82}{84}
\emline{21.59}{16.82}{85}{23.09}{16.28}{86}
\emline{23.09}{16.28}{87}{24.43}{15.42}{88}
\emline{24.43}{15.42}{89}{25.54}{14.27}{90}
\emline{25.54}{14.27}{91}{26.37}{12.91}{92}
\emline{26.37}{12.91}{93}{26.86}{11.39}{94}
\emline{26.86}{11.39}{95}{27.00}{9.80}{96}
\emline{27.00}{9.80}{97}{26.77}{8.22}{98}
\emline{26.77}{8.22}{99}{26.19}{6.73}{100}
\emline{26.19}{6.73}{101}{25.29}{5.42}{102}
\emline{25.29}{5.42}{103}{24.11}{4.34}{104}
\emline{24.11}{4.34}{105}{22.72}{3.55}{106}
\emline{22.72}{3.55}{107}{21.19}{3.10}{108}
\emline{21.19}{3.10}{109}{19.60}{3.01}{110}
\emline{19.60}{3.01}{111}{18.03}{3.28}{112}
\emline{18.03}{3.28}{113}{16.56}{3.90}{114}
\emline{16.56}{3.90}{115}{15.27}{4.84}{116}
\emline{15.27}{4.84}{117}{14.22}{6.05}{118}
\emline{14.22}{6.05}{119}{13.48}{7.46}{120}
\emline{13.48}{7.46}{121}{13.07}{9.00}{122}
\emline{13.07}{9.00}{123}{13.03}{10.60}{124}
\emline{13.03}{10.60}{125}{13.34}{12.16}{126}
\emline{13.34}{12.16}{127}{14.01}{13.61}{128}
\emline{14.01}{13.61}{129}{14.98}{14.88}{130}
\emline{14.98}{14.88}{131}{16.22}{15.89}{132}
\emline{16.22}{15.89}{133}{17.65}{16.59}{134}
\emline{17.65}{16.59}{135}{20.00}{17.00}{136}
%\end
\put(20.00,10.00){\circle{10.00}}
\put(50.00,10.00){\circle{10.00}}
%\bezvec{80}(46.00,7.00)(36.33,4.33)(26.33,7.00)
\put(26.33,7.00){\vector(-4,1){0.2}}
\emline{46.00}{7.00}{137}{43.58}{6.42}{138}
\emline{43.58}{6.42}{139}{41.15}{6.00}{140}
\emline{41.15}{6.00}{141}{38.70}{5.75}{142}
\emline{38.70}{5.75}{143}{33.79}{5.75}{144}
\emline{33.79}{5.75}{145}{31.31}{6.00}{146}
\emline{31.31}{6.00}{147}{28.83}{6.42}{148}
\emline{28.83}{6.42}{149}{26.33}{7.00}{150}
%\end
%\bezvec{80}(26.33,13.00)(37.00,15.67)(46.00,13.00)
\put(46.00,13.00){\vector(4,-1){0.2}}
\emline{26.33}{13.00}{151}{28.97}{13.58}{152}
\emline{28.97}{13.58}{153}{31.56}{14.00}{154}
\emline{31.56}{14.00}{155}{34.10}{14.25}{156}
\emline{34.10}{14.25}{157}{39.02}{14.25}{158}
\emline{39.02}{14.25}{159}{41.40}{14.00}{160}
\emline{41.40}{14.00}{161}{43.72}{13.58}{162}
\emline{43.72}{13.58}{163}{46.00}{13.00}{164}
%\end
\put(20.00,10.00){\makebox(0,0)[cc]{$q_3$}}
\put(50.00,10.00){\makebox(0,0)[cc]{$q_2$}}
%\bezvec{132}(54.00,13.00)(70.00,10.00)(54.00,7.00)
\put(54.00,7.00){\vector(-4,-1){0.2}}
\emline{54.00}{13.00}{165}{56.24}{12.55}{166}
\emline{56.24}{12.55}{167}{58.11}{12.09}{168}
\emline{58.11}{12.09}{169}{59.62}{11.64}{170}
\emline{59.62}{11.64}{171}{60.76}{11.18}{172}
\emline{60.76}{11.18}{173}{61.53}{10.73}{174}
\emline{61.53}{10.73}{175}{61.93}{10.27}{176}
\emline{61.93}{10.27}{177}{61.97}{9.82}{178}
\emline{61.97}{9.82}{179}{61.64}{9.36}{180}
\emline{61.64}{9.36}{181}{60.94}{8.91}{182}
\emline{60.94}{8.91}{183}{59.88}{8.45}{184}
\emline{59.88}{8.45}{185}{58.44}{8.00}{186}
\emline{58.44}{8.00}{187}{56.64}{7.55}{188}
\emline{56.64}{7.55}{189}{54.00}{7.00}{190}
%\end
%\bezvec{116}(13.67,13.00)(-0.33,10.00)(13.67,7.00)
\put(13.67,7.00){\vector(4,-1){0.2}}
\emline{13.67}{13.00}{191}{11.46}{12.48}{192}
\emline{11.46}{12.48}{193}{9.67}{11.97}{194}
\emline{9.67}{11.97}{195}{8.30}{11.45}{196}
\emline{8.30}{11.45}{197}{7.34}{10.93}{198}
\emline{7.34}{10.93}{199}{6.80}{10.41}{200}
\emline{6.80}{10.41}{201}{6.67}{9.90}{202}
\emline{6.67}{9.90}{203}{6.97}{9.38}{204}
\emline{6.97}{9.38}{205}{7.67}{8.86}{206}
\emline{7.67}{8.86}{207}{8.80}{8.34}{208}
\emline{8.80}{8.34}{209}{10.34}{7.83}{210}
\emline{10.34}{7.83}{211}{13.67}{7.00}{212}
%\end
\put(5.00,25.00){\circle{10.00}}
\put(5.00,25.00){\makebox(0,0)[cc]{$q_5$}}
%\bezvec{132}(9.00,28.00)(25.00,25.00)(9.00,22.00)
\put(9.00,22.00){\vector(-4,-1){0.2}}
\emline{9.00}{28.00}{213}{11.24}{27.55}{214}
\emline{11.24}{27.55}{215}{13.11}{27.09}{216}
\emline{13.11}{27.09}{217}{14.62}{26.64}{218}
\emline{14.62}{26.64}{219}{15.76}{26.18}{220}
\emline{15.76}{26.18}{221}{16.53}{25.73}{222}
\emline{16.53}{25.73}{223}{16.93}{25.27}{224}
\emline{16.93}{25.27}{225}{16.97}{24.82}{226}
\emline{16.97}{24.82}{227}{16.64}{24.36}{228}
\emline{16.64}{24.36}{229}{15.94}{23.91}{230}
\emline{15.94}{23.91}{231}{14.88}{23.45}{232}
\emline{14.88}{23.45}{233}{13.44}{23.00}{234}
\emline{13.44}{23.00}{235}{11.64}{22.55}{236}
\emline{11.64}{22.55}{237}{9.00}{22.00}{238}
%\end
\put(12.67,30.00){\makebox(0,0)[cc]{a,b}}
%\bezvec{84}(53.00,33.67)(57.00,24.00)(53.00,14.00)
\put(53.00,14.00){\vector(-1,-3){0.2}}
\emline{53.00}{33.67}{239}{53.84}{31.36}{240}
\emline{53.84}{31.36}{241}{54.45}{29.04}{242}
\emline{54.45}{29.04}{243}{54.84}{26.72}{244}
\emline{54.84}{26.72}{245}{55.00}{24.38}{246}
\emline{55.00}{24.38}{247}{54.93}{22.04}{248}
\emline{54.93}{22.04}{249}{54.63}{19.69}{250}
\emline{54.63}{19.69}{251}{54.11}{17.32}{252}
\emline{54.11}{17.32}{253}{53.00}{14.00}{254}
%\end
\put(6.00,40.00){\makebox(0,0)[cc]{$b$}}
\put(35.00,47.00){\makebox(0,0)[cc]{$a$}}
\put(35.00,39.00){\makebox(0,0)[cc]{$a$}}
\put(57.00,25.00){\makebox(0,0)[cc]{$b$}}
\put(58.00,15.00){\makebox(0,0)[cc]{$b$}}
\put(35.00,17.00){\makebox(0,0)[cc]{$a$}}
\put(35.00,8.00){\makebox(0,0)[cc]{$a$}}
\put(10.00,14.67){\makebox(0,0)[cc]{$b$}}
%\bezvec{72}(15.00,40.00)(6.00,39.00)(5.00,30.00)
\put(5.00,30.00){\vector(-1,-4){0.2}}
\emline{15.00}{40.00}{255}{12.65}{39.57}{256}
\emline{12.65}{39.57}{257}{10.62}{38.83}{258}
\emline{10.62}{38.83}{259}{8.89}{37.78}{260}
\emline{8.89}{37.78}{261}{7.47}{36.42}{262}
\emline{7.47}{36.42}{263}{6.36}{34.75}{264}
\emline{6.36}{34.75}{265}{5.56}{32.78}{266}
\emline{5.56}{32.78}{267}{5.00}{30.00}{268}
%\end
%\vector(66.33,43.00)(56.33,43.00)
\put(56.33,43.00){\vector(-1,0){0.2}}
\emline{66.33}{43.00}{269}{56.33}{43.00}{270}
%\end
\end{picture}
    \end{Large}
    \setlength{\abovecaptionskip}{0pt}
    \caption{Automaton $G_3$}
    \label{Bilde4}
  \end{minipage}
\end{figure}
%***************************************************

So, the automaton $K_2$ will consist of 8 states:
$q_1$, $q_2$, $q_3$, $q_4$, $q_5$, $q_6$, $q_7$, $q_8$, where
$Q_{non}=\{q_1, q_2, q_3, q_4\}$ , $Q_{acc}=\{q_5, q_8\}$ , $Q_{rej}=\{q_6, q_7\}$.

The unitary transform matrixes $V_\kappa$, $V_a$, $V_b$ and $V_\$$ are:
$$V_\kappa=
\left (
\begin{array}{cccccccc}
\sqrt\frac{2}{3}&\sqrt\frac{1}{3}&0&0&0&0&0&0\\
\sqrt\frac{1}{3}&-\sqrt\frac{2}{3}&0&0&0&0&0&0\\
0&0&-\sqrt\frac{2}{3}&\sqrt\frac{1}{3}&0&0&0&0\\
0&0&\sqrt\frac{2}{3}&\sqrt\frac{1}{3}&0&0&0&0\\
0&0&0&0&1&0&0&0\\
0&0&0&0&0&1&0&0\\
0&0&0&0&0&0&1&0\\
0&0&0&0&0&0&0&1
\end{array}
\right ),
V_a=
\left (
\begin{array}{cccccccc}
0&0&0&1&0&0&0&0\\
0&0&1&0&0&0&0&0\\
0&1&0&0&0&0&0&0\\
1&0&0&0&0&0&0&0\\
0&0&0&0&1&0&0&0\\
0&0&0&0&0&1&0&0\\
0&0&0&0&0&0&1&0\\
0&0&0&0&0&0&0&1
\end{array}
\right )
$$

$$V_b=
\left (
\begin{array}{cccccccc}
0&0&0&0&\sqrt\frac{1}{2}&\sqrt\frac{1}{2}&0&0\\
0&1&0&0&0&0&0&0\\
0&0&1&0&0&0&0&0\\
0&0&0&0&0&0&1&0\\
\sqrt\frac{1}{2}&0&0&0&\frac{1}{2}&-\frac{1}{2}&0&0\\
\sqrt\frac{1}{2}&0&0&0&-\frac{1}{2}&\frac{1}{2}&0&0\\
0&0&0&1&0&0&0&0\\
0&0&0&0&0&0&0&1
\end{array}
\right ),
V_\$=
\left (
\begin{array}{cccccccc}
0&0&0&0&1&0&0&0\\
0&0&0&0&0&1&0&0\\
0&0&0&0&0&0&0&1\\
0&0&0&0&0&0&1&0\\
1&0&0&0&0&0&0&0\\
0&1&0&0&0&0&0&0\\
0&0&0&1&0&0&0&0\\
0&0&1&0&0&0&0&0
\end{array}
\right )
$$

The starting state for $K_2$ is $q_1$,
for $K_3$ it is $q_4$.
Now we will look only at $K_2$. For $K_3$ it is similar.

State $q_1$ in $G_2$ corresponds to 
$\psi_1=\sqrt\frac{2}{3}\ket{q_1}+\sqrt\frac{1}{3}\ket{q_2}$ in $K_2$

State $q_2$ in $G_2$ corresponds to 
$\psi_2=\sqrt\frac{1}{3}\ket{q_2}$ in $K_2$

State $q_3$ in $G_2$ corresponds to 
$\psi_3=\sqrt\frac{1}{3}\ket{q_3}$ in $K_2$

State $q_4$ in $G_2$ corresponds to 
$\psi_4=\sqrt\frac{2}{3}\ket{q_4}+\sqrt\frac{1}{3}\ket{q_3}$ in $K_2$

%Easy to see that
%
%\begin{enumerate}
%\item
%After reading the left endmarker $\kappa$ automaton is in state $\psi_1$
%or $V_\kappa(\ket{q_1})=\psi_1$, also starting state of $G_2$ is $q_1$
%\item
%If by reading letter a or b automaton $G_2$ passes from $q_i$ to $q_j$
%then automaton $K_2$ passes from $\psi_i$ to $\psi_j$.
%\item
%If automaton $K_3$ is in state $q_4$ and receives letter $b$
%then it rejects input with probability $\frac{2}{3}$ so we have
%no speial interest what happens further.
%\item
%When automaton $G_2$ is in state $q_2$ or $q_3$, then with probability $\frac{1}{3}$
%it has been accepted input, with probability $\frac{1}{3}$ rejected,
%or it is in corresponding state $\psi_2$ or $\psi_3$.
%\item
%If automaton receives the right endmarker in state $\psi_1$
%then input is accepted with probability $\frac{2}{3}$
%\item
%If automaton receives the right endmarker in state $\psi_2$
%then input is rejected with probability $\frac{1}{3}$
%and as it was rejected with same probability so far,
%the total probability to reject input is $\frac{2}{3}$
%\item
%If automaton receives the right endmarker in state $\psi_3$
%%and as it was accepted with same probability so far,
%the total probability to reject input is $\frac{2}{3}$
%\item
%If automaton receives the right endmarker in state $\psi_4$
%then input is rejected with probability $\frac{2}{3}$
%\end{enumerate}
%
%I think this must be enough to proove, that automaton $K_2$
%recognizes language $L_2$ and $K_3$ recognizes $L_3$
%with probability $\frac{2}{3}$
%
\begin{enumerate}
\item
After reading the left endmarker $\kappa$ automaton is in state $\psi_1$
or $V_\kappa'(\ket{q_1})=\psi_1$, also starting state of $G_2$ is $q_1$.
\item
If by reading letter $a$ automaton $G_2$ passes from $q_1$ to $q_4$
or back, then automaton $K_2$ state changes from $\psi_1$ to $\psi_4$
or back.
\item
If automaton $K_3$ is in state $q_4$ and receives letter $b$
then it rejects input with probability $\frac{2}{3}$ so we have
no special interest what happens further (and it is correct,
because $G_2$ is now in "all rejecting" state $q_5$).
\item
If automaton $G_2$ is in state $q_1$ and receives letter $b$
it passes to $q_3$, if automaton $K_2$ is in state $\psi_1$
and receives letter $b$ it passes to state
$\frac{1}{\sqrt{3}}\ket{q_2}+\frac{1}{\sqrt{3}}\ket{q_5}+\frac{1}{\sqrt{3}}\ket{q_6}$
and after measurement accepts input with
probability $\frac{1}{3}$,
rejects input with the same probability $\frac{1}{3}$,
or continues in state $\psi_2$.
\item
If by reading letter $a$ automaton $G_2$ passes from $q_2$ to $q_3$
or back, then automaton's $K_2$ state changes from $\psi_2$ to $\psi_3$
or back. By reading letter $b$ $G_2$ passes from
$q_2$ to $q_2$ and from $q_3$ to $q_3$. Also $K_2$ -
if it is in $\psi_2$ or $\psi_3$ and receives $b$ it
does not change its state.
\item
If automaton receives the right endmarker in state $\psi_1$
then input is accepted with probability $\frac{2}{3}$.
\item
If automaton receives the right endmarker in state $\psi_2$
then input is rejected with probability $\frac{1}{3}$
and as it was rejected with same probability so far,
the total probability to reject input is $\frac{2}{3}$.
\item
If automaton receives the right endmarker in state $\psi_3$
then input is accepted with probability $\frac{1}{3}$
and as it was accepted with same probability so far,
the total probability to reject input is $\frac{2}{3}$.
\item
If automaton receives the right endmarker in state $\psi_4$
then input is rejected with probability $\frac{2}{3}$.
\end{enumerate}

In these 9 points we wanted to show, that automaton
$K_2$ performs computation the same way as $G_1$.
While automaton $G_2$ is in one of its states
$q_1,\dots  ,q_4$, $K$ is in a corresponding state
$\psi_1,\dots , \psi_4$.
Automaton$K_2$ accepts input with probability
$\frac{2}{3}$
iff it receives right endmarker $\$$ in one of
states $\psi_1$ or $\psi_3$, corresponding whom
$q_1$ and $q_3$ are the only accepting states in $G_1$.
So we can conclude,
that $K_2$ accepts language $L_2$ with
probability $\frac{2}{3}$.

%Now we have described, how automaton $K_2$ (and the same way $K_3$)
%works. This is also a proof, that automaton $K_2$
%recognizes language $L_2$ and $K_3$ recognizes $L_3$
%with probability $\frac{2}{3}$.

What are languages $L_1$, $L_2$ and $L_3$ informally?

$L_3$ consists of all words which start with \underline{odd} number of
letters $a$ and after first letter $b$ (if there is such) there
is odd number of letters $a$.

$L_2$ consists of all words which start with \underline{even} number of
letters $a$ and after first letter $b$ (if there is such) there
is odd number of letters $a$.

$L_1$ consists of all words which start with \underline{any} number of
letters $a$ and after first letter $b$ (if there is such) there
is odd number of letters $a$.

So, it is almost evident, that $L_1=L_2\bigcup L_3$.

\begin{corollary}
\label{cor1}
There are two languages $L_2$ and $L_3$ which are recognizable by QFA
(with probability $\frac{2}{3}$),
union of whom $L_1=L_2\bigcup L_3$ is not recognizable by QFA.
\end{corollary}

\begin{corollary}
\label{cor2}
The class of languages recognizable by QFA is not closed under union.
\end{corollary}

As $L_2\bigcap L_3=\emptyset$ then also $L_1=L_2 \Delta L_3$.
So the class of languages recognizable by QFA
is not closed under symmetric difference.
From this and from the fact, that this class
is closed under complement easy follows:

\begin{corollary}
\label{cor3}
The class of languages recognizable by QFA is not closed under any binary
Boolean operation, where both arguments are significant.
\end{corollary}

\section{Some more details}

In previous section we found two languages $L_2$ and $L_3$
recognizable by QFA with probability $\frac{2}{3}$,
union of whom is not recognizable by any QFA.
What if we increase the probability?

\begin{theorem}
\label{T31}
If 2 languages $L_1$ and $L_2$ are recognizable by QFA with
probabilities $p_1$ and $p_2$ and $\frac{1}{p_1}+\frac{1}{p_2}<3$,
then $L=L_1\bigcup L_2$ is also recognizable by QFA with
probability $\frac{2p_1p_2}{p_1+p_2+p_1p_2}$.

In case if $p_1, p_2>\frac{2}{3}$ the condition holds.
\end{theorem}

\begin{proof}
We have automaton $K_1$, which accepts $L_1$ with probability $p_1$ and
automaton $K_2$, which accepts $L_2$ with probability $p_2$.
We will make automaton $K$ which will work like this:
\begin{enumerate}
\item
Runs $K_1$ with probability $\frac{p_2}{p_1+p_2+p_1p_2}$,
\item
Runs $K_2$ with probability $\frac{p_1}{p_1+p_2+p_1p_2}$,
\item
Accepts input with probability $\frac{p_1p_2}{p_1+p_2+p_1p_2}$.
\end{enumerate}

To make such an automaton we just have to make tensor product $K_1\otimes K_2\otimes K_3$
where $K_3$ consists of only one "all accepting" state, and modify a little
its $V_\kappa$ matrix. When we have done it, we have:

\begin{enumerate}
\item
$w\cin L_1$ and $w\cin L_2\longrightarrow$ input is accepted
with probability

$$\frac{p_2}{p_1+p_2+p_1p_2}*p_1+\frac{p_1}{p_1+p_2+p_1p_2}*p_2+
\frac{p_1p_2}{p_1+p_2+p_1p_2}*1=1$$
\item
$w\cin L_1$ and $w\cnotin L_2\longrightarrow$ input is accepted
with probability at least

$$\frac{p_2}{p_1+p_2+p_1p_2}*p_1+\frac{p_1p_2}{p_1+p_2+p_1p_2}*1=
\frac{2p_1p_2}{p_1+p_2+p_1p_2}$$
\item
$w\cnotin L_1$ and $w\cin L_2\longrightarrow$ input is accepted
with probability at least

$$\frac{p_1}{p_1+p_2+p_1p_2}*p_2+\frac{p_1p_2}{p_1+p_2+p_1p_2}*1=
\frac{2p_1p_2}{p_1+p_2+p_1p_2}$$
\item
$w\cnotin L_1$ and $w\cnotin L_2\longrightarrow$ input is rejected
with probability at least

$$\frac{p_2}{p_1+p_2+p_1p_2}*p_1+\frac{p_1}{p_1+p_2+p_1p_2}*p_2=
\frac{2p_1p_2}{p_1+p_2+p_1p_2}$$
%\qed
\end{enumerate}
So automaton $K$ recognizes $L$ with probability
at least
$$\frac{2p_1p_2}{p_1+p_2+p_1p_2}=
\frac{1}{2}+\frac{3-(\frac{1}{p_1}+
\frac{1}{p_2})}{4(1+\frac{1}{p_1}+\frac{1}{p_2})}>\frac{1}{2}$$
\end{proof}

All this has also a nice geometric interpretation.
We are going to build a linear
function $f$ from probabilities $x_1$, $x_2$
to probability $x$ such, that
$f(p_1,p_2)\geq\frac{1}{2}+\varepsilon$,
$f(p_1,0)\geq\frac{1}{2}+\varepsilon$, 
$f(0,p_2)\geq\frac{1}{2}+\varepsilon$, 
$f(1-p_1,1-p_2)\leq\frac{1}{2}-\varepsilon$.
Geometrically we consider a plane $x,y$ where each word $w$ is
located in a point $(x,y)$, where $x$ is probability
that $K_1$ accepts $w$ and $y$ is the probability,
that $K_2$ accepts $w$.

%**************************************
\begin{figure}
  \setlength{\abovecaptionskip}{0pt}
  \begin{minipage}{0.5\linewidth}
    \centering
%TexCad Options
%\grade{\on}
%\emlines{\on}
%\beziermacro{\off}
%\reduce{\on}
%\snapping{\off}
%\quality{2.00}
%\graddiff{0.01}
%\snapasp{1}
%\zoom{1.00}
\special{em:linewidth 0.4pt}
\unitlength 1.00mm
\linethickness{0.4pt}
\begin{picture}(50.00,45.00)
%\vector(10.00,5.00)(50.00,5.00)
\put(50.00,5.00){\vector(1,0){0.2}}
\emline{10.00}{5.00}{1}{50.00}{5.00}{2}
%\end
%\vector(10.00,5.00)(10.00,45.00)
\put(10.00,45.00){\vector(0,1){0.2}}
\emline{10.00}{5.00}{3}{10.00}{45.00}{4}
%\end
\emline{16.00}{5.00}{5}{16.00}{17.00}{6}
\emline{16.00}{17.00}{7}{10.00}{17.00}{8}
\emline{34.00}{5.00}{9}{34.00}{17.00}{10}
\emline{40.00}{17.00}{11}{40.00}{5.00}{12}
\emline{10.00}{23.00}{13}{16.00}{23.00}{14}
\emline{16.00}{35.00}{15}{10.00}{35.00}{16}
\emline{40.00}{23.00}{17}{40.00}{35.00}{18}
\emline{40.00}{17.00}{19}{40.00}{23.00}{20}
\emline{34.00}{17.00}{21}{34.00}{23.00}{22}
\emline{34.00}{23.00}{23}{16.00}{23.00}{24}
\emline{34.00}{35.00}{25}{40.00}{35.00}{26}
\emline{34.00}{35.00}{27}{16.00}{35.00}{28}
\begin{large}
\put(13.00,11.00){\makebox(0,0)[cc]{$S_1$}}
\put(37.00,29.00){\makebox(0,0)[cc]{$S_2$}}
\end{large}
\emline{-1.33}{31.00}{29}{45.00}{-3.67}{30}
\put(7.00,2.00){\makebox(0,0)[cc]{$0$}}
\put(16.00,2.00){\makebox(0,0)[cc]{$1\!-\!p_1$}}
\put(34.00,2.00){\makebox(0,0)[cc]{$p_1$}}
\put(40.00,2.00){\makebox(0,0)[cc]{$1$}}
\put(49.00,2.00){\makebox(0,0)[cc]{$x$}}
\put(7.00,17.00){\makebox(0,0)[rc]{$1\!-\!p_2\!\!$}}
\put(7.00,23.00){\makebox(0,0)[cc]{$p_2$}}
\put(7.00,35.33){\makebox(0,0)[cc]{$1$}}
\put(7.00,44.00){\makebox(0,0)[cc]{$y$}}
\end{picture}
    \setlength{\abovecaptionskip}{0pt}
    \setlength{\belowcaptionskip}{30pt}
    \caption{}
    \label{Bilde5}
  \end{minipage}%
  \begin{minipage}{0.5\linewidth}
    \centering
%TexCad Options
%\grade{\on}
%\emlines{\on}
%\beziermacro{\off}
%\reduce{\on}
%\snapping{\off}
%\quality{2.00}
%\graddiff{0.01}
%\snapasp{1}
%\zoom{1.00}
\special{em:linewidth 0.4pt}
\unitlength 1.00mm
\linethickness{0.4pt}
\begin{picture}(50.00,45.00)
%\vector(10.00,5.00)(50.00,5.00)
\put(50.00,5.00){\vector(1,0){0.2}}
\emline{10.00}{5.00}{1}{50.00}{5.00}{2}
%\end
%\vector(10.00,5.00)(10.00,45.00)
\put(10.00,45.00){\vector(0,1){0.2}}
\emline{10.00}{5.00}{3}{10.00}{45.00}{4}
%\end
\emline{40.00}{17.00}{5}{40.00}{5.00}{6}
\emline{16.00}{35.00}{7}{10.00}{35.00}{8}
\emline{40.00}{23.00}{9}{40.00}{35.00}{10}
\emline{40.00}{17.00}{11}{40.00}{23.00}{12}
\emline{34.00}{35.00}{13}{40.00}{35.00}{14}
\emline{34.00}{35.00}{15}{16.00}{35.00}{16}
\begin{large}
\put(16.00,11.00){\makebox(0,0)[cc]{$S_1$}}
\put(33.00,28.00){\makebox(0,0)[cc]{$S_2$}}
\end{large}
\put(7.00,2.00){\makebox(0,0)[cc]{$0$}}
\put(22.00,2.00){\makebox(0,0)[cc]{$1\!-\!p_1$}}
\put(28.00,2.00){\makebox(0,0)[cc]{$p_1$}}
\put(40.00,2.00){\makebox(0,0)[cc]{$1$}}
\put(49.00,2.00){\makebox(0,0)[cc]{$x$}}
\put(7.00,17.00){\makebox(0,0)[rc]{$1\!-\!p_2\!\!$}}
\put(7.00,23.00){\makebox(0,0)[cc]{$p_2$}}
\put(7.00,35.33){\makebox(0,0)[cc]{$1$}}
\put(7.00,44.00){\makebox(0,0)[cc]{$y$}}
\emline{22.00}{5.00}{17}{22.00}{17.00}{18}
\emline{22.00}{17.00}{19}{10.00}{17.00}{20}
\emline{28.00}{5.00}{21}{28.00}{23.00}{22}
\emline{28.00}{23.00}{23}{10.00}{23.00}{24}
\emline{5.00}{30.00}{25}{38.00}{-3.00}{26}
\end{picture}
    \setlength{\abovecaptionskip}{0pt}
    \setlength{\belowcaptionskip}{30pt}
    \caption{}
    \label{Bilde6}
  \end{minipage}\\
  \begin{minipage}{0.5\linewidth}
    \centering
%TexCad Options
%\grade{\on}
%\emlines{\on}
%\beziermacro{\off}
%\reduce{\on}
%\snapping{\off}
%\quality{2.00}
%\graddiff{0.01}
%\snapasp{1}
%\zoom{1.00}
\special{em:linewidth 0.4pt}
\unitlength 1.00mm
\linethickness{0.4pt}
\begin{picture}(50.00,45.00)
%\vector(10.00,5.00)(50.00,5.00)
\put(50.00,5.00){\vector(1,0){0.2}}
\emline{10.00}{5.00}{1}{50.00}{5.00}{2}
%\end
%\vector(10.00,5.00)(10.00,45.00)
\put(10.00,45.00){\vector(0,1){0.2}}
\emline{10.00}{5.00}{3}{10.00}{45.00}{4}
%\end
\emline{40.00}{17.00}{5}{40.00}{5.00}{6}
\emline{16.00}{35.00}{7}{10.00}{35.00}{8}
\emline{40.00}{23.00}{9}{40.00}{35.00}{10}
\emline{40.00}{17.00}{11}{40.00}{23.00}{12}
\emline{34.00}{35.00}{13}{40.00}{35.00}{14}
\emline{34.00}{35.00}{15}{16.00}{35.00}{16}
\begin{large}
\put(15.00,10.00){\makebox(0,0)[cc]{$S_1$}}
\put(34.00,29.00){\makebox(0,0)[cc]{$S_2$}}
\end{large}
\put(7.00,2.00){\makebox(0,0)[cc]{$0$}}
\put(7.00,35.33){\makebox(0,0)[cc]{$1$}}
\put(7.00,44.00){\makebox(0,0)[cc]{$y$}}
\emline{20.00}{5.00}{17}{20.00}{15.00}{18}
\emline{20.00}{15.00}{19}{10.00}{15.00}{20}
\emline{30.00}{5.00}{21}{30.00}{25.00}{22}
\emline{30.00}{25.00}{23}{10.00}{25.00}{24}
\emline{7.00}{28.00}{25}{35.00}{0.00}{26}
\put(20.33,2.00){\makebox(0,0)[cc]{$\frac{1}{3}$}}
\put(30.33,2.00){\makebox(0,0)[cc]{$\frac{2}{3}$}}
\put(40.33,2.00){\makebox(0,0)[cc]{$1$}}
\put(7.33,14.00){\makebox(0,0)[cc]{$\frac{1}{3}$}}
\put(7.33,24.00){\makebox(0,0)[cc]{$\frac{2}{3}$}}
\end{picture}
    \setlength{\abovecaptionskip}{0pt}
    \setlength{\belowcaptionskip}{30pt}
    \caption{}
    \label{Bilde9}
  \end{minipage}%
  \begin{minipage}{0.5\linewidth}
    \centering
%TexCad Options
%\grade{\on}
%\emlines{\on}
%\beziermacro{\off}
%\reduce{\on}
%\snapping{\off}
%\quality{2.00}
%\graddiff{0.01}
%\snapasp{1}
%\zoom{1.00}
\special{em:linewidth 0.4pt}
\unitlength 1.00mm
\linethickness{0.4pt}
\begin{picture}(50.00,45.00)
%\vector(10.00,5.00)(50.00,5.00)
\put(50.00,5.00){\vector(1,0){0.2}}
\emline{10.00}{5.00}{1}{50.00}{5.00}{2}
%\end
%\vector(10.00,5.00)(10.00,45.00)
\put(10.00,45.00){\vector(0,1){0.2}}
\emline{10.00}{5.00}{3}{10.00}{45.00}{4}
%\end
\begin{large}
\put(20.00,15.00){\makebox(0,0)[cc]{$S_1$}}
\put(30.00,25.00){\makebox(0,0)[cc]{$S_2$}}
\end{large}
\put(7.00,2.00){\makebox(0,0)[cc]{$0$}}
\put(22.00,2.00){\makebox(0,0)[cc]{$1\!-\!p_1$}}
\put(28.00,2.00){\makebox(0,0)[cc]{$p_1$}}
\put(40.00,2.00){\makebox(0,0)[cc]{$1$}}
\put(49.00,2.00){\makebox(0,0)[cc]{$x$}}
\put(7.00,17.00){\makebox(0,0)[rc]{$1\!-\!p_2\!\!$}}
\put(7.00,23.00){\makebox(0,0)[cc]{$p_2$}}
\put(7.00,35.33){\makebox(0,0)[cc]{$1$}}
\put(7.00,44.00){\makebox(0,0)[cc]{$y$}}
\emline{22.00}{17.00}{5}{22.00}{13.00}{6}
\emline{22.00}{13.00}{7}{18.00}{13.00}{8}
\emline{18.00}{13.00}{9}{18.00}{17.00}{10}
\emline{18.00}{17.00}{11}{22.00}{17.00}{12}
\emline{28.00}{13.00}{13}{32.00}{13.00}{14}
\emline{32.00}{13.00}{15}{32.00}{27.00}{16}
\emline{32.00}{27.00}{17}{18.00}{27.00}{18}
\emline{18.00}{27.00}{19}{18.00}{23.00}{20}
\emline{18.00}{23.00}{21}{28.00}{23.00}{22}
\emline{28.00}{23.00}{23}{28.00}{13.00}{24}
\emline{8.00}{32.00}{25}{41.00}{0.00}{26}
\end{picture}
    \setlength{\abovecaptionskip}{0pt}
    \setlength{\belowcaptionskip}{30pt}
    \caption{}
    \label{Bilde7}
  \end{minipage}\\
  \centering
%TexCad Options
%\grade{\on}
%\emlines{\on}
%\beziermacro{\off}
%\reduce{\on}
%\snapping{\off}
%\quality{2.00}
%\graddiff{0.01}
%\snapasp{1}
%\zoom{1.00}
\special{em:linewidth 0.4pt}
\unitlength 1.00mm
\linethickness{0.4pt}
\begin{picture}(50.00,45.00)
%\vector(10.00,5.00)(50.00,5.00)
\put(50.00,5.00){\vector(1,0){0.2}}
\emline{10.00}{5.00}{1}{50.00}{5.00}{2}
%\end
%\vector(10.00,5.00)(10.00,45.00)
\put(10.00,45.00){\vector(0,1){0.2}}
\emline{10.00}{5.00}{3}{10.00}{45.00}{4}
%\end
\begin{large}
\put(20.00,15.00){\makebox(0,0)[cc]{$S_1$}}
\put(30.00,25.00){\makebox(0,0)[cc]{$S_2$}}
\end{large}
\put(7.00,2.00){\makebox(0,0)[cc]{$0$}}
\put(22.00,2.00){\makebox(0,0)[cc]{$1\!-\!p_1$}}
\put(28.00,2.00){\makebox(0,0)[cc]{$p_1$}}
\put(40.00,2.00){\makebox(0,0)[cc]{$1$}}
\put(49.00,2.00){\makebox(0,0)[cc]{$x$}}
\put(7.00,17.00){\makebox(0,0)[rc]{$1\!-\!p_2\!\!$}}
\put(7.00,23.00){\makebox(0,0)[cc]{$p_2$}}
\put(7.00,35.33){\makebox(0,0)[cc]{$1$}}
\put(7.00,44.00){\makebox(0,0)[cc]{$y$}}
\emline{9.00}{32.67}{5}{42.00}{0.67}{6}
\emline{40.00}{4.00}{7}{40.00}{6.00}{8}
\emline{9.00}{35.00}{9}{11.00}{35.00}{10}
\emline{28.00}{4.00}{11}{28.00}{6.00}{12}
\emline{22.00}{4.00}{13}{22.00}{6.00}{14}
\emline{9.00}{17.00}{15}{11.00}{17.00}{16}
\emline{9.00}{23.00}{17}{11.00}{23.00}{18}
\emline{28.00}{17.00}{19}{28.00}{23.00}{20}
\emline{28.00}{23.00}{21}{22.00}{23.00}{22}
\put(22.00,17.00){\circle*{0.67}}
\end{picture}
  \setlength{\abovecaptionskip}{0pt}
  \setlength{\belowcaptionskip}{30pt}
  \caption{}
  \label{Bilde8}
\end{figure}
%**************************************

$S_1$ is the place, where lies all words, that do not belong to $L$.

$S_2$ is the place, where lies all words, that belong to $L$.

If we can (Fig.\ref{Bilde5}) separate these two parts with a line $ax+by=c$
then we can construct automaton "$K=aK_1+bK_2$"
with $c$ as isolated cut point. 
If we can not (Fig.\ref{Bilde6}), then this method doesn't help.
And as it was shown higher, sometimes none of other methods can help, too.

Case when $p_1=p_2=\frac{2}{3}$ (Fig.\ref{Bilde9}) is the limit case.
If any of the probabilities were a little bit greater
then this method would help.

Sometimes it may be, that there are no words $w$ such, that
$K_1$ or $K_2$ would reject with probability $1-t$ or greater.
Then (Fig.\ref{Bilde7}) you can see, that now it is easier, to make
such a line, so condition $\frac{1}{p_1}+\frac{1}{p_2}<3$
can be weakened
(the probabilities in Fig.\ref{Bilde7}
are the same as in Fig.\ref{Bilde6}).
In the limit case, when rejecting probabilities are only
$p_1$ and $p_2$, $S_1$ is the point $(1-p_1,1-p_2)$ (Fig.\ref{Bilde8}).
So with any $p_1$ and $p_2$ you can separate $S_1$ from $S_2$
with a line, from what follows you can always
construct $K=K_1\bigcup K_2$.

Now it is clear, that languages $L_2$ and $L_3$ defined in
chapter 2, cannot be recognized with probability greater than $\frac{2}{3}$
so the construction presented there is best possible.

\section{Appendix - proof of theorem \ref{T12}}
%\repeat{T12}
In this proof we are going to use one classical result
from \cite{BV 97}, so as it has very little connection with
all other proof, we are going to present it here, in the beginning.
\begin{lemma}
\label{LBV}
If $\psi$ and $\phi$ are two states of quantum system and
$\|\psi-\phi\|<\varepsilon$ then total variation distance
between probability distributions generated by
measurements on $\psi$ and $\phi$ are less than $2\varepsilon$.
\end{lemma}
\begin{proof}
Let's denote
$$\varphi=\frac{1}{2}(\psi+\phi)=\sum_i\alpha_i\ket{q_i}$$
and
$$\pi=\frac{1}{2}(\psi-\phi)=\sum_i\gamma_i\ket{q_i}\;,\;
\|\pi\|<\frac{\varepsilon}{2}$$
The total variation distance between two probability
distributions $P=\sum p_i\ket{q_i}$
and $R=\sum r_i\ket{q_i}$ is defined as
$$\Delta=\sum_i|p_i-r_i|$$
As $\psi=\varphi+\pi$ and $\phi=\varphi-\pi$ then
total variation distance is
$$\Delta=\sum_i|\|\alpha_i+\gamma_i\|^2-\|\alpha_i-\gamma_i\|^2|=$$
$$=\sum_i|2\alpha_i\gamma_i^*+2\alpha_i^*\gamma_i|\leq
4\sum_i|\alpha_i||\gamma_i|$$
%Now let's look at two vectors
%$\varphi'=\Sigma_i|\alpha_i|\ket{q_i}$ and
%$\pi'=\Sigma_i|\gamma_i|\ket{q_i}$.
%Their coordinates are modulos of coordinates
%of $\varphi$ and $\pi$, so their length also is
%the length of $\varphi$ and $\pi$. But their scalar product
%$\langle\pi'\ket{\varphi'}=\Sigma_i|\alpha_i||\gamma_i|$,
%so $\Delta\leq 4\langle\pi'\ket{\varphi'}\leq
%4\|\varphi\|\|\pi\|=2\|\varphi\|\varepsilon$
Now using Cauchy inequality we get
$$\Delta\leq 4\sqrt{\sum_i|\alpha_i|^2*\sum_i|\gamma_i|^2}=
4\|\varphi\|\|\pi\|<2\|\varphi\|\varepsilon$$ and as
$\|\varphi\|\leq 1$ then $\Delta<2\varepsilon$.
\end{proof}

This lemma shows the intuitively clear fact, that
close states are accepted with close probabilities.
In our proof we are going to use it in such a form, that
difference between acceptance probabilities (and also
rejection probabilities ) of states $\psi$ and $\phi$
where $\|\psi-\phi\|<\varepsilon$
is less than $2\varepsilon$.

Let's say, that there is such QFA $K$, which recognizes
the same language as $G$ with a fixed probability $\frac{1}{2}+\varepsilon$.
First thing we will do is decompose
it´s state space $E_{non}$ into 2 components $E_{non}=E_1\oplus E_2$.
In $E_1$ we will put all vectors $\psi$ with such a property:
if automaton $K$ starts in $\psi$ then the probability, that
input is accepted or rejected while reading any word $w\cin \Sigma^*$
is $0$. Or $\forall w\cin\Sigma^*\; \|\psi\|=\|V_w'(\psi)\|$.
$E_2$ will contain all vectors orthogonal to $E_1$.
\begin{tabbing}
More formally \=\+ we will do it this way:\\
$E^0=E_{non}$\\
$E^1=\{\psi\;|\;\psi\cin E^0 \; \& \; V_a(\psi)\cin E^0 \; 
\& \; V_b(\psi)\cin E^0\}$\\
$E^2=\{\psi\;|\;\psi\cin E^1 \; \& \; V_a(\psi)\cin E^1 \;
\& \; V_b(\psi)\cin E^1\}$\\
$E^3=\{\psi\;|\;\psi\cin E^2 \; \& \; V_a(\psi)\cin E^2 \;
\& \; V_b(\psi)\cin E^2\}$\\
$\dots$\\
$E^{j+1}=\{\psi\;|\;\psi\cin E^j \; \& \; V_a(\psi)\cin E^j \;
\& \; V_b(\psi)\cin E^j\}$\\
$E_1=\bigcap_{j=0}^{+\infty}E^j\;\;\;\;\;\;\;\;\;E_2=E\ominus E$\\
\end{tabbing}

At first we will notice, that $E^{j+1}\!\!\subseteq E^j$,
so $\dim E^{j+1}\!\!\leq \dim E^j$ .
If $\dim E^j=\dim E^{j+1}$ then $E^j=E^{j+1}=E^{j+2}=...$, hence
$\forall j\!\geq\!n\; E^j=E^n$, where $n=\dim E_{non}$, or
$n$ is just the number of states in $Q_{non}$.
So as well we can define $E_1=\bigcap_{j=0}^{n}E^j$.
This means, that for each state $\psi$ not in $E_1$ there is a word of length $n$,
which projects part of $\psi$ to $Q_{acc}$ or $Q_{rej}$.
As in $E_1$ there are no projections, then 
$V_a'(\psi)=V_a(\psi)$ and $V_b'(\psi)=V_b(\psi)$ if $\psi\cin E_1$, so 
$V_a'$ and $V_b'$ are unitary in $E_1$. And as product of 2 unitary matrixes
is also unitary, so $V_w'$ is unitary in $E_1$
for all $w\cin\Sigma^*$.
From the definition of $E_1$ follows, that
$\forall\psi\cin E_1\Rightarrow V_a'(\psi)\cin E_1$ and $V_b'(\psi)\cin E_1$.
As unitary transformations transforms orthogonal vectors to orthogonal,
we can conclude, that
$$\forall\psi\cin E_2\Rightarrow V_a(\psi)\cin E_2\oplus E_{acc}\oplus E_{rej},
 V_b(\psi)\cin E_2\oplus E_{acc}\oplus E_{rej}$$  therefore
$$\forall\psi\cin E_2\Rightarrow V_a'(\psi)\cin E_2 , V_b'(\psi)\cin E_2$$
So we can say, that computation is performed in $E_1$ and $E_2$ independently.

\begin{lemma}
\label{L41}
For every $\psi\cin E_2$ and every $\delta$
there is such a word $w\cin\Sigma^*$, that
$\|V_w'(\psi)\|<\delta$ or in other words
$\inf\{\|V_w'(\psi)\|\;|\; \psi\cin E_2,w\cin\Sigma^*\}=0$.
\end{lemma}
\begin{proof}
For each vector $\psi\cin E_2$ let's denote
$M_\psi=min\{\|V_w'(\psi)\|\; |\: w\cin\Sigma^n\}$
and $M=\{M_\psi\;|\; \psi\cin E_2,\|\psi\|\leq 1\}$
where $n$ is still the number of states 
in $Q_{non}$. It means, that for each $\psi$ we
find a word $w$ with length $n$ reading which
automaton would make maximum projections.
It is clear, that $M_\psi<1$, otherwise $\psi$ would be in $E_1$.
We denote $S=\sup(M)$.
As set $\{\psi\; |\; \psi\cin E_2\;\|\psi\|\leq 1\}$
is closed, so is $M$. Hence $S\cin M$ and so $S<1$.
Now the proof is easy. For each $\psi\cin E_2$
we can construct word $w\cin\Sigma^{kn}$
such, that $\|V_w'(\psi)\|\leq S^k\|\psi\|
\rightarrow 0$ when $k\rightarrow\infty$.
\end{proof}

We'll say, that state $\psi_1$ is reachable from state $\psi_2$,
if there is a sequence of words $\{w_i\}$ such, that
$$\lim_{i\to\infty}\|V_{w_i}'(\psi_2)-\psi_1\| = 0$$
Let's put $\delta_i=\|V_{w_i}'(\psi_2)-\psi_1\|$, now
$\delta_i\to 0$, when $i\to\infty$.
Let's look at sequence of vectors
$$\psi_1, U(\psi_1), U^2(\psi_1), U^3(\psi_1),\dots$$ where $U=V_{w_i}'$.
As all they are inside finite space???, and they are infinitely many,
then I can find a pair of them as close to one another as I wish,
say $$\|U^k(\psi_1)-U^m(\psi_1)\|<\delta_i ,\; k<m$$
Then $$\|U^k(\psi_1-U^{m-k}(\psi_1))\|<\delta_i$$ and also
$$\|\psi_1-U^{m-k}(\psi_1)\|<\delta_i$$
because unitary transformation doesn't change
the length of vector. So now we have $\|U(\psi_2)-\psi_1\|=\delta_i$
and $\|\psi_1-U^{m-k}(\psi_1)\|<\delta_i$. By triangle
inequality we can conclude, that
$$\|U(\psi_2)-U^{m-k}(\psi_1)\|<2\delta_i$$
or $\|\psi_2-U^{m-k-1}(\psi_1)\|<2\delta_i$.
What does it mean? If we denote $u_i=w_i^{m-k-1}$
($m$ and $k$ may be different for each $w_i$) then
$$\lim_{i\to\infty}\|V_{u_i}'(\psi_1)-\psi_2\|
\leq \lim_{i\to\infty}2\delta_i=0$$
or reachability is symmetric.

It is also very easy to prove, that reachability is transitive.
It follows directly from the fact, that
transformations are continuous.

%Let's take $\psi_1$ reachable from $\psi_2$, and
%$\psi_2$ reachable from $\psi_3$. It means, that we have 2 sequences
%of words $u_i$ and $w_i$, that
%$\|V_{u_i}'(\psi_2)-\psi_1\|=\varepsilon_i\to 0$ and
%$\|V_{w_i}'(\psi_3)-\psi_2\|=\delta_i\to 0$ when $i\to\infty$.
%We take the sequence $w_iu_i$.
%$\|V_{w_iu_i}'(\psi_3)-\psi_1\|\leq
%\|V_{w_iu_i}'(\psi_3)-V_{u_i}'(\psi_2)\|+\|V_{u_i}'(\psi_2)-\psi_1\|<
%\delta_i+\varepsilon_i\to 0$ when $i\to\infty$

To prove the transitivity of reachability
we even did not need the unitarity of transformations,
we used only their continuity, so reachability is
transitive in $E_{non}$, and symmetric in $E_1$,
where the transformations are unitary.
So it is equivalence in $E_1$.

Let's denote the state after reading left endmarker
$\psi_0=\psi_I+\psi_{II}$, where $\psi_I\cin E_1$ and $\psi_{II}\cin E_2$.
Also after reading any word $w\cin \Sigma^*$, the state is
$V_w'(\psi_0)=V_w'(\psi_I)+V_w'(\psi_{II})$,
where $V_w'(\psi_I)\cin E_1$ and $V_w'(\psi_{II})\cin E_2$.
Let's denote $R$ the class of all reachable
states from starting state $\psi_I$.
Also let's denote $A(\psi)$ the probability
to accept input, if automaton in state
$\psi$ receives right endmarker $\$$,
and $p_w$ the probability, that it has accepted input, while reading
word $\kappa w$. So the probability that automaton accepts
word $w$ is $p_w + A(\psi_w)$.

We begin with reading word $w$ such, that $\|V_w'(\psi_{II})\|<k$, where $k$ is very small.
We can easily assume, that automaton $G_1$ after reading $w$ is in state $q_2$, if it is not,
then instead of $w$ we can take $wb$ or $wa$ if it is in $q_1$ or $q_3$.

$$\psi_w=V_w'(\psi_0)=V_w'(\psi_I)+V_w'(\psi_{II})=\psi_w^1+\psi_w^2 ,\; \|\psi_w^2\|<k$$
In further calculation we can omit existence of $\psi_w^2$,
and assume, that $\psi_w=\psi_w^1$, and $\forall u\cin\Sigma^*\; p_w\!=\!p_{wu}$
because probability changes $\psi_2$ can make, are too small, when the
difference between acceptance and rejection probabilities has to
be at least $2\varepsilon$.
%As $\psi_w^2$ is the only part, that can be projected, if we now read any word $u$,
%then

%$p_{wu}\leq p_w+\|\psi_w^2\|^2<p_w+k^2$\\
%Now we'll divide $R$ into 3 subsets.\\
%$R_{1k=}\{\psi | \frac{1}{2}+\varepsilon-8k-k^2\leq p_w+A(\psi)\leq 1\}$\\
%$R_{2k=}\{\psi | \frac{1}{2}-\varepsilon+8k< p_w+A(\psi)<\frac{1}{2}+\varepsilon-8k-k^2\}$\\
%$R_{3k=}\{\psi | 0\leq p_w+A(\psi)\leq\frac{1}{2}-\varepsilon+8k\}$\\

%$R_{2k}$ is empty

%Let there be $\psi\cin R_{2k}$. As it is reachable from $\psi_w^1$, then
%there is word $u$, that $\|V_u'(\psi_w^1)-\psi\|<k$,we also have
%$\|V_u'(\psi_w^2)\|\leq\|\psi_w^2)\|< k$, what by
%triangle inequality implies
%$\|\psi-V_u'(\psi_w)\|<2k$.
%Then by lemma XXX $|A(\psi)-A(\psi_{wu})|<8k$.
%So as $\frac{1}{2}-\varepsilon+8k\leq p_w+A(\psi)\leq\frac{1}{2}+\varepsilon-8k-k^2$
%then $\frac{1}{2}-\varepsilon< p_{wu}+A(\psi_{wu})<\frac{1}{2}+\varepsilon$,
%so the automaton accepts word $wu$ with probability between
%$\frac{1}{2}-\varepsilon$ and $\frac{1}{2}+\varepsilon$ - contradiction.

Now we will divide $R$ into 3 subsets.\\
$R_1=\{\psi \;|\; \frac{1}{2}+\varepsilon\leq p_w+A(\psi)\leq 1\}$\\
$R_2=\{\psi \;|\; \frac{1}{2}-\varepsilon< p_w+A(\psi)<\frac{1}{2}+\varepsilon\}$\\
$R_3=\{\psi \;|\; 0\leq p_w+A(\psi)\leq\frac{1}{2}-\varepsilon\}$\\
\begin{lemma}
\label{L42}
$R_2$ is empty.
\end{lemma}
\begin{proof}
Let there be $\psi\cin R_2$, we denote
$max(\frac{1}{2}+\varepsilon-\|\psi\|,\|\psi\|-\frac{1}{2}+\varepsilon)=2k$.
As $\psi$ is reachable from $\psi_w$, then
there is word $u$, that $\|V_u'(\psi_w)-\psi\|<k$.

Then by lemma \ref{LBV} $|A(\psi)-A(\psi_{wu})|<2k$.
So as $\frac{1}{2}-\varepsilon+2k\leq p_w+A(\psi)\leq\frac{1}{2}
+\varepsilon-2k$
then $\frac{1}{2}-\varepsilon< p_{wu}+A(\psi_{wu})
<\frac{1}{2}+\varepsilon$,
so the automaton accepts word $wu$ with probability between
$\frac{1}{2}-\varepsilon$ and $\frac{1}{2}+\varepsilon$ - contradiction.
\end{proof}
If automaton is in state $\psi\cin R_1$ and receives right endmarker $\$$,
it accepts input.
If automaton is in state $\psi\cin R_3$ and receives right endmarker $\$$,
it rejects input.

After reading letter $a$ automaton must change its state from state, where
it accepts input ($R_1$) to state, where it doesn't accept it ($R_3$),
and vice versa, reading of letter $b$ should not change anything.
More formally
$$\forall\psi\cin R_1\Rightarrow V_a'(\psi)\cin R_3, V_b'(\psi)\cin R_1$$
$$\forall\psi\cin R_3\Rightarrow V_a'(\psi)\cin R_1, V_b'(\psi)\cin R_3$$

Now we have 2 choices:
\begin{enumerate}
\item
$\psi_I\cin R_1$. Let's look at states $\psi_{bw}$ and $\psi_{bwa}$,
where word $w$ is chosen, to make $V_{bw}'(\psi_{II})$ negligible,
and contains pair number of $a$-s (we can always find such).
From this our choice follows, that
$\psi_{bw}\cin R_1$ and $\psi_{bwa}\cin R_3$, so probability
to accept word $bw$ is greater
than probability to accept $bwa$, 
at least for $2\varepsilon$
what is not correct, because
$bwa$ belongs to language
but $bw$ does not.
\item
$\psi_I\cin R_3$. The same problem.
Let's look at states $\psi_{abw}$ and $\psi_{abwa}$,
where word $w$ is chosen, to make $V_{abw}'(\psi_{II})$ negligible,
and contains pair number of $a$-s (we can always find such).
From this our choice follows, that
$\psi_{abw}\cin R_1$ and $\psi_{abwa}\cin R_3$, so probability
to accept word $abw$ is greater
than probability to accept $abwa$, 
at least for $2\varepsilon$
what is not correct, because
$abwa$ belongs to language
but $abw$ does not.
\end{enumerate}

So we have found, that automaton $K$ does not recognize some
words correctly, so it does not recognize language $L_1$.
Now the proof is finished.

\end{document}